Jianxin Sun · David Lenz · Hongfeng Yu · Tom Peterka

# MFA-DVR: direct volume rendering of MFA models

**Abstract** 3D volume rendering is widely used to reveal insightful intrinsic patterns of volumetric datasets across many domains. However, the complex structures and varying scales of volumetric data can make efficiently generating high-quality volume rendering results a challenging task. Multivariate functional approximation (MFA) is a new data model that addresses some of the critical challenges: high-order evaluation of both value and derivative anywhere in the spatial domain, compact representation for large-scale volumetric data, and uniform representation of both structured and unstructured data. In this paper, we present MFA-DVR, the first direct volume rendering pipeline utilizing the MFA model, for both structured and unstructured volumetric datasets. We demonstrate improved rendering quality using MFA-DVR on both synthetic and real datasets through a comparative study. We show that MFA-DVR not only generates more faithful volume rendering than using local filters but also performs faster on high-order interpolations on structured and unstructured datasets. MFA-DVR is implemented in the existing volume rendering pipeline of the Visualization Toolkit (VTK) to be accessible by the scientific visualization community.

## 1 Introduction

Volume rendering is a popular and effective visualization technique to analyze volumetric datasets from various scientific domains, such as medical imaging, meteorology, materials science, and physical simulations. However, volume renderings can produce artifacts caused by inaccurate value and gradient interpolation on samples of interest within the volume. Rendering speed is another crucial factor for real-time applications and interactive visualization. Different volume rendering algorithms are often needed for structured and unstructured data, further complicating their use.

Volume visualization algorithms such as ray casting require accurate querying of values and gradients at arbitrary locations within the 3D volumetric space to generate high-quality images. Volumetric datasets are usually represented in discrete regular grids or irregular meshes. For such discrete volumetric datasets, the values and gradients of points at arbitrary locations are generally calculated locally from neighboring samples through interpolation from voxels or faces. The most straightforward interpolation method used in popular volume visualization applications and tools is trilinear interpolation (Mroz et al. 2000) which has been optimized for performance through bitwise operations and hardware acceleration (Csébfalvi 2019). However, such interpolation is first-order, which can produce aliasing effects and high-frequency artifacts for sparse data, small objects (Krylov et al. 2009), or complex internal structures. Although other high-older local filters can mitigate high-frequency artifacts, they can still give less correct interpolation results due to overfitting on specific data samples. Gradients are even more sensitive to the quality of the interpolation when used for illumination (Moller et al. 1997).

Cell size, shape, and unstructured mesh connectivity make unstructured volume rendering more complicated and slower than its structured counterpart (Moreland and Angel 2004). Ray casting-based unstructured volume rendering algorithms involve expensive searches for mesh primitives to calculate the interpolated value of the intersection between the ray and primitive. Cell projection algorithms rely on visibility ordering (Comba et al. 1999; Callahan et al. 2005) of polyhedra to optimize the rendering process. Although GPUs can speed up the volume rendering of unstructured data, rendering complex unstructured data of large size remain a challenge.

Multivariate functional approximation (MFA) (Peterka et al. 2018) is a compact high-order multivariate continuous representation of a dataset that enables directly querying for more accurate values and gradients at any location on demand without interpolating nearby samples. In this paper, we propose an MFA-based volume rendering pipeline based on MFA (MFA-DVR) with improved rendering quality and moderate time complexity. This paper is the first study of directly rendering the MFA model for volume visualization. MFA-DVR can eliminate linear interpolation artifacts on complex datasets, providing higher accuracy and rendering faster than local high-order filters. MFA-DVR is also capable of rendering unstructured data faster with higher quality than traditional unstructured volume rendering algorithms. We quantitatively and qualitatively evaluate the rendering quality and performance of MFA-DVR through comparative studies with other popular structured and unstructured volume rendering filters and algorithms. We explore how the internal parameters of MFA affect the rendering quality and performance. We implemented MFA-DVR in the Visualization Toolkit (VTK) (Schroeder et al. 1996) for its accessibility to the community. The source code is available online. The contributions of this paper are:

- The first volume rendering pipeline (MFA-DVR) directly rendering the MFA model with high rendering quality and moderate time complexity.
- Implementation and optimization of querying the MFA model for both value and gradient efficiently.
- Experiments studying the behavior and performance of MFA-DVR with a number of datasets, rendering scenarios, and configurations.

The remainder of this paper is structured as follows: we first introduce related work in Sect. 2. Section 3 details the architecture and implementation of the proposed MFA-DVR. Experimental results and evaluation are presented in Sect. 4. We provide conclusions in Sect. 5.

## 2 Related work

In the following, we review previous works that share similarities with our approach from the fields of high-order filters for interpolation and volume visualization.

### 2.1 Interpolation artifacts

Increasing the resolution of a dataset can mitigate zero- or first-order interpolation artifacts, but it can be difficult to acquire high-resolution data in situ for domains (Zhang et al. 2010). 3D super-resolution and upscaling approaches (Shilling et al. 2008; Gholipour et al. 2010) can be used to increase resolution. Recent works utilize artificial neural networks for super-resolution in 3D volumetric data for spatial (Weiss et al.

2022; Guo et al. 2020), temporal (Han and Wang 2022; Han and Wang 2019), and spatiotemporal domains (Han et al. 2021; An et al. 2021) to improve the quality of volume and isosurface rendering. However, super-resolution methods increase data size, demanding a more extensive storage and memory footprint. The increased size can also slow down the rendering time, which is critical for interactive volume visualization. Moreover, linear interpolation artifacts can still exist and become noticeable when the zoom level approaches the resolution of the upscaled dataset.

Linear interpolation artifacts can be overcome by using high-order interpolation filters. Existing works replace trilinear interpolation with tricubic (Lekien and Marsden 2005; Sigg and Hadwiger 2005), cubic B-spline (Ruijters et al. 2008; Lee et al. 2010; Vuçini et al. 2009) and Catmull-Rom (Csébfalvi 2018) filters for better accuracy. Original discrete data needs to be pre-processed first through encoding or prefiltering (Ruijters and Thévenaz 2012; Thévenaz et al. 2000) to obtain the sequence of coefficients of the interpolating functions. Rendering is achieved by querying the sample points on the ray segment from the constructed interpolating functions. Using such higher-order interpolation avoids generating larger datasets, as super-resolution methods do, at the cost of increased computational complexity for querying. Such high-order interpolation filters need to be carefully treated for the best rendering result. Filters with infinite spatial extent, such as Gaussian and sinc, require their intrinsic parameters and the type of windowing function to be tuned to prevent severe ringing and postaliasing artifacts (Theußl et al. 2000). Such complicated steps hinder high-order interpolation methods from easily being used in popular volume visualization tools.

Our proposed solution is similar to high-order interpolation to mitigate low-order interpolation artifacts for complex datasets. However, MFA is a global approximation rather than a local fitting, so MFA-DVR gives better global accuracy compared with other high-order local filters. Also, the MFA model of the entire dataset is encoded a priori rather than during rendering, allowing MFA-DVR to render faster than high-order local filters.

### 2.2 Unstructured volume visualization

Current DVR algorithms on structured data are highly optimized, utilizing GPU acceleration and ray traversal optimization such as space-leaping (Cho et al. 1999). In comparison, volume rendering of unstructured data is more challenging and mainly requires expensive search operations on unstructured meshes with complex geometry. When rendering point clouds, a mesh structure needs to be constructed using triangulation, which can be slow for large datasets. Much research has been done over the past decades on optimizing various aspects of volume rendering on unstructured data. GPU shaders (Klein et al. 2004; Wylie et al. 2002; Muigg et al. 2011), CUDA (Okuyan et al. 2014; Larsen et al. 2015; Maximo et al. 2008) and ray tracing (RT) (Morrical et al. 2020, 2019; Şahıstan et al. 2021) have been leveraged to accelerate rendering speed. Memory efficiency and data transfer latency between system memory and GPU memory have been optimized through efficient data structures and partitions for parallelism (Sahistan et al. 2022; Wald et al. 2021; Marmitt and Slusallek 2006) when rendering large-scale data. Even with GPU acceleration and parallelism, unstructured volume rendering is generally slower than structured volume rendering under similar constraints (Silva et al. 2005). Moreover, due to the complexity of the geometry of unstructured data, rendering quality is not only affected by interpolation artifacts (Kraus et al. 2004) but can also be distorted by variations in mesh resolution and structure.

The same functional representation of the MFA model for both structured and unstructured data allows the same ray casting algorithm to be used regardless of the nature of the original raw data. Because the MFA is a continuous meshless representation, the proposed MFA-DVR involves no triangulation of input data points to create unstructured meshes needed by many unstructured volume rendering algorithms, and it avoids complicated searching and projecting operations of such unstructured meshes.

### 2.3 MFA

The MFA (Peterka et al. 2018; Nashed et al. 2019) models discrete high-dimensional scientific data by a functional basis representation based on a tensor product of nonuniform rational B-spline functions. Both field geometric information and values of the scientific data are modeled and efficiently represented by a set of control points and knots. Figure 1 demonstrates the pipeline of fitting an MFA model from raw input data. Initially, parameterization and initial knot distribution are computed for the minimum number of control points. New control points and knots are added adaptively until all evaluated points in each span of knots are within the allowable maximum relative error of the original points. Throughout this iterative procedure, knot spans that fall beyond the tolerance are subdivided, and MFA is recomputed.

Compared with other local filters, which try to solve a large number of small local optimization problems, MFA is the solution of a single global optimization over the entire domain. MFA has several advantages over local filters: First, local filters interpolate or fit exactly to input points, whereas MFA approximates or fits approximately to the input points. This allows smoothing of high-frequency discretization artifacts while capturing underlying low-frequency phenomena. Second, local filters result in reduced continuity between adjacent filter invocations. MFA, being a global model, has full high-order continuity everywhere. Third, the size and location of local filters are governed by the input point distribution, being a sliding window over the input points. In contrast, the piecewise polynomials in the MFA are distributed according to the knot positions which are independent of the input point distribution. The knot locations can reflect the complexity of the data, rather than the input point distribution. Finally, local filters may rely on finite approximations for the gradient and produce approximate high-order derivatives, especially at cell boundaries, whereas MFA provides analytical high-order derivatives. While the model is an approximation, its derived quantities such as gradients are analytically correct. Higher-order derivatives, up to the polynomial degree, are also analytically available.

Although MFA is capable of modeling large-scale scientific data in situ, there is no existing volume visualization pipeline prior to this paper that visualizes an encoded MFA model directly to support analysis post hoc. In this work, our focus is on how to efficiently query the MFA model for both value and gradient at arbitrary spatial locations within the volume in the context of volume rending. Additionally, we conducted a comprehensive assessment of the quality and performance of directly rendering an MFA model, a novel investigation not previously undertaken. Furthermore, our research presents a user-friendly solution for integrating the MFA model into the popular VTK library's pipeline, enabling researchers from various scientific disciplines to explore this new model and improve the visualization of their volumetric data. MFA-DVR leverages the above features of MFA for high global accuracy across the volumetric space, which is crucial to generate faithful volume visualization. Sections 4.2 and 4.3 will demonstrate that MFA-DVR not only outperforms trilinear interpolation on rendering quality, but also popular high-order filters like tricubic and Catmull-Rom on both accuracy and performance.

## 3 Volume rendering of the MFA model

Figure 2 provides the architecture of the MFA-DVR pipeline. At some point prior to rendering, the raw volumetric dataset, which can be structured or unstructured, is encoded into the MFA model. This encoding process can be configured by polynomial degrees and a number of control points. Details can be found in (Peterka et al. 2018). Since the encoding process is not the focus of this paper, we assume that the encoded MFA mode is available as the input of our renderer. The MFA model has multiple uses besides volume rendering, and we assume that early in the workflow, discrete data are converted into the MFA model, which is then used for subsequent processing, analysis and visualization. The MFA-DVR pipeline takes the encoded MFA model and directly generates the final volume rendering. Among the blocks of the MFA-DVR pipeline, MFA value query and gradient query blocks are optimized for querying value and gradient at samples on the ray directly from the MFA model efficiently with high-order accuracy. In traditional volume rendering algorithms, such values and gradients are often interpolated using local filters on neighboring samples either on vertices of voxels for structured datasets, or from vertices of mesh elements for unstructured datasets.

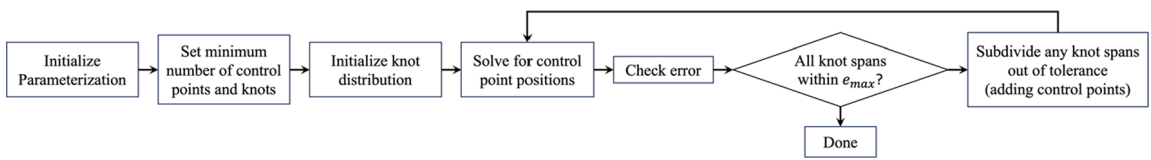

**Fig. 1** An overall of MFA encoding algorithm, where $e_{\max}$ is the allowable maximum relative error

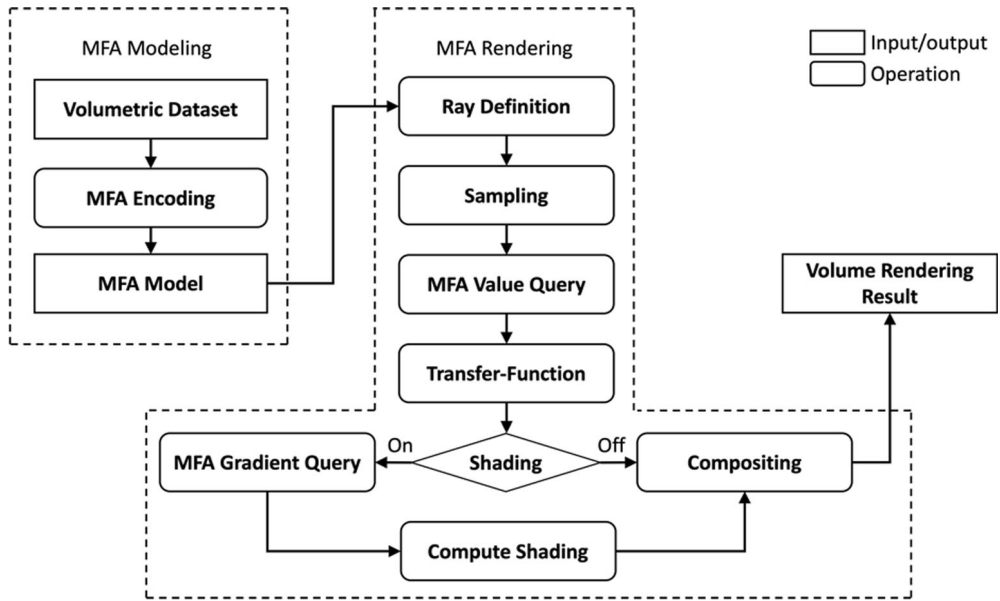

**Fig. 2** Proposed MFA-DVR volume rendering architecture

3.1 MFA rendering

The encoded MFA model is a surrogate data model representing the input data and facilitating the retrieval of values and gradients at arbitrary locations anywhere in the domain directly from the model. This feature distinguishes MFA from other representations such as wavelets, cosine transformations, and compression algorithms that require the inverse transform to be applied. The values and gradients queried from the MFA model are exact, analytical, high order, and closed-form with higher accuracy than local filters and finite difference estimation.

The proposed MFA-DVR is a ray casting-based volume renderer with high rendering quality and parallelizable architecture. Ray casting is a popular and widely implemented method that casts virtual light rays through each pixel of the resulting image viewport and accumulates samples on those rays by following various models (Max 1995) of light interaction through compositing. The fundamental issue is how to effectively and efficiently provide value and gradient from the MFA model for the points of interest used for color compositing, which has not been studied before.

Ray Definition is the first step where the fixed-point ray cast model was used to define the ray cast for each pixel of the viewport according to the relative location of the camera and the viewport. Although samples on the ray in this work are selected with constant sample distance, dynamic adaptive sampling algorithms (Wang et al. 2020; Hachisuka et al. 2008) can also leverage our MFA rendering method for better rendering speed and sampling efficiency. For each sample, the value is retrieved by calling the MFA value query function from the MFA model and then mapped to opacity and color according to transfer functions. If shading is disabled, a front-to-back traversal compositing process accumulates the color and opacity to calculate the final values of color and opacity (RGBA) for the pixel of the rendered image. If shading is enabled, the gradient at each sample also needs to be retrieved by calling MFA first-order derivative query function from the MFA model. Diffuse and specular coefficients are computed by taking the gradients together with the current view direction, light direction, ambient color, diffuse color, specular color, light intensity and material parameters. Color and opacity values with shading will then be calculated with diffuse and specular coefficients, and later used for the same compositing process for the final RGBA results of the specific pixel. We minimized the querying overhead by optimizing the value query and computing all derivatives along the three dimensions at once to speed up the ray traversal. Higher-order derivatives can also be queried from the MFA model as needed for a wider range of applications, like volume feature detection (Persoon et al. 2003) and curvature-based transfer function (Kindlmann et al. 2003). To accelerate the rendering process, early ray termination (ERT) is used to stop the composition near the end of ray traversal, and computation for all pixel values of the viewport can also be parallelized.

**Algorithm 1** MFA rendering steps
---
**Input:** MFA model.
**Output:** RGBA value for pixels allocated for each thread.
1: **for** $i \leftarrow Row_{ThreadStart}$ to $Row_{ThreadEnd}$ **do**
2:     **for** $j \leftarrow Colume_{LeftBounds}$ to $Colume_{RightBounds}$ **do**
3:         **for** $s \leftarrow Sample_{First}$ to $Sample_{Last}$ **do**
4:             $x, y, z = getSampleLocation(s)$
5:             $x, y, z = normalizeSampleLocation(x, y, z)$
6:             $v = queryValueMFA(x, y, z)$
7:             $applyTFs(TF_{Color}, TF_{Opacity}, v, color)$
8:             **if** $ShadingOn == ture$ **then**
9:                 $g_x, g_t, g_z = queryGradientMFA(x, y, z)$
10:                $c_{def}, c_{spe} = getLightingCoefficient(g_x, g_t, g_z)$
11:                $addShading(c_{def}, c_{spe}, color)$
12:             **end if**
13:             $O_{cumulative} = doColorCompositing(color)$
14:             **if** $O_{cumulative} > O_{Max}$ **then**    ▷Early termination
15:                break
16:             **end if**
17:         **end for**
18:         $setPixelColor(color)$
19:     **end for**
20: **end for**

The pseudocode of MFA rendering is presented in Algorithm 1. Each thread has access to the MFA model and is in charge of processing rays from pixels of different rows of the viewport. The outer for-loop, from lines 1 to 20, assigns specific rows of the viewport to be processed by the current thread. The middle for-loop, from lines 2 to 19, selects pixels of each assigned row within the bounds, from left to right, of the volume object and delivers the final pixel color. The inner loop, from lines 3 to 17, processes ray traversal and compositing for each pixel. $Sample_{First}$ is the first sample on the ray inside the volume as the ray enters the volume, while the $Sample_{Last}$ is the last sample on the ray inside the volume as the ray leaves the volume. Before querying the MFA model, the 3D coordinates of the sample are normalized to the range [0, 1] to match the parameterization of the MFA model. Querying functions for value and gradient can be called as needed through our efficient MFA-DVR querying interfaces queryValueMFA() and queryGradientMFA() with minimized latency. applyTFs() applies color transfer function $TF_{color}$ and opacity transfer function $TF_{Opacity}$ and updates color. getLightingCoefficient() computes the diffuse coefficients $c_{def}$ and specular coefficients $c_{spe}$. addShading() computes the RGBA values with shading and updates color. doColorCompositing() composites and updates *color* for the final pixel assigned by setPixelColor() and returns the current cumulative opacity $O_{cumulative}$ which will be used for checking early termination by comparing with maximal opacity $O_{max}$.

The rendering process of MFA-DVR has several advantages over traditional ray casting volume rendering algorithm: First, for shading, MFA-DVR directly queries gradient values from the MFA model on demand for shading computation, so no gradient table, which is three times (three-dimensional derivatives for each sample) the size of the input data, is allocated and computed in advance. Such a feature saves system memory footprint when rendering large-scale datasets and makes the rendering application starts faster. Second, MFA-DVR does not create and maintain the shading table, storing lighting coefficients according to the current camera parameters, whenever the user's view changes. This gives MFA-DVR a better response time for interactive visualization. Third, MFA-DVR can render the MFA model regardless of the original data structure. Both structured and unstructured data can be rendered accurately and efficiently through the same MFA-DVR pipeline. Although some algorithms are capable of calculating gradient and shading parameters on demand through local filters, their computational complexity can cause performance degradation. Section 4.3 will show a detailed evaluation of performance through a comparison between MFA-DVR and other first or high-order filters.

### 3.2 Implementation in VTK

MFA-DVR is implemented into the popular Visualization Toolkit (VTK), an open-source foundation for visualization work worldwide, with straightforward interfaces. We hope to eventually provide MFA-DVR as a VTK plugin for others to use. Existing volume visualization tools and applications using the VTK library require minimal changes to render structured or unstructured data using MFA-DVR. Only the MFA model is needed and passed by reference to the vtkAlgorithm class through its child class vtkFixedPointVolumeRayCastMapper by calling our MFA interface function SetMFAInputConnection(), making the model accessible from the MFA-DVR pipeline. Figure 3 shows such data flow and rendering call chain from the existing VTK volume rendering stack down to our MFA-DVR rendering code. We follow the same naming convention used by VTK and implement MFA-DVR functions for volume rendering with and without shading in this work.

## 4 Results and evaluation

We use both synthetic and real datasets in structured and unstructured formats to evaluate MFA-DVR rendering quality and performance. We also investigate how the key MFA parameters affect such aspects. Since there are extensive existing works on rendering structure and unstructured datasets with various data structure, optimization, and hardware acceleration strategies, in this work, we only select several important volume rendering algorithms with typical types of value and gradient interpolation methods. For structured datasets, we considered widely used trilinear interpolation first-order filter, and tricubic and Catmull-Rom high-order filters. The implementation of tricubic is based on Lekien (Lekien and Marsden 2005), which gives state-of-the-art tricubic performance. We also integrate such high-order filters into the pipeline of rendering in VTK. For unstructured datasets, we selected the popular Bunyk (Bunyk et al. 1997) and Zsweep (Farias et al. 2000) as two representatives of ray casting-based unstructured volume rendering algorithms. We also considered Projected Tetrahedra (PT) (Shirley and Tuchman 1990), which is a typical implementation of view-independent cell projection (Weiler et al. 2003) utilizing projective techniques (Moreland and Angel 2004). PT is also the building block of many hardware-accelerated volume rendering algorithms (Silva et al. 2005; Beyer et al. 2014). The VTK implementation of such unstructured volume rendering algorithms is used in this study. All algorithms considered are evaluated under the same visualization pipeline of VTK for a fair comparison.

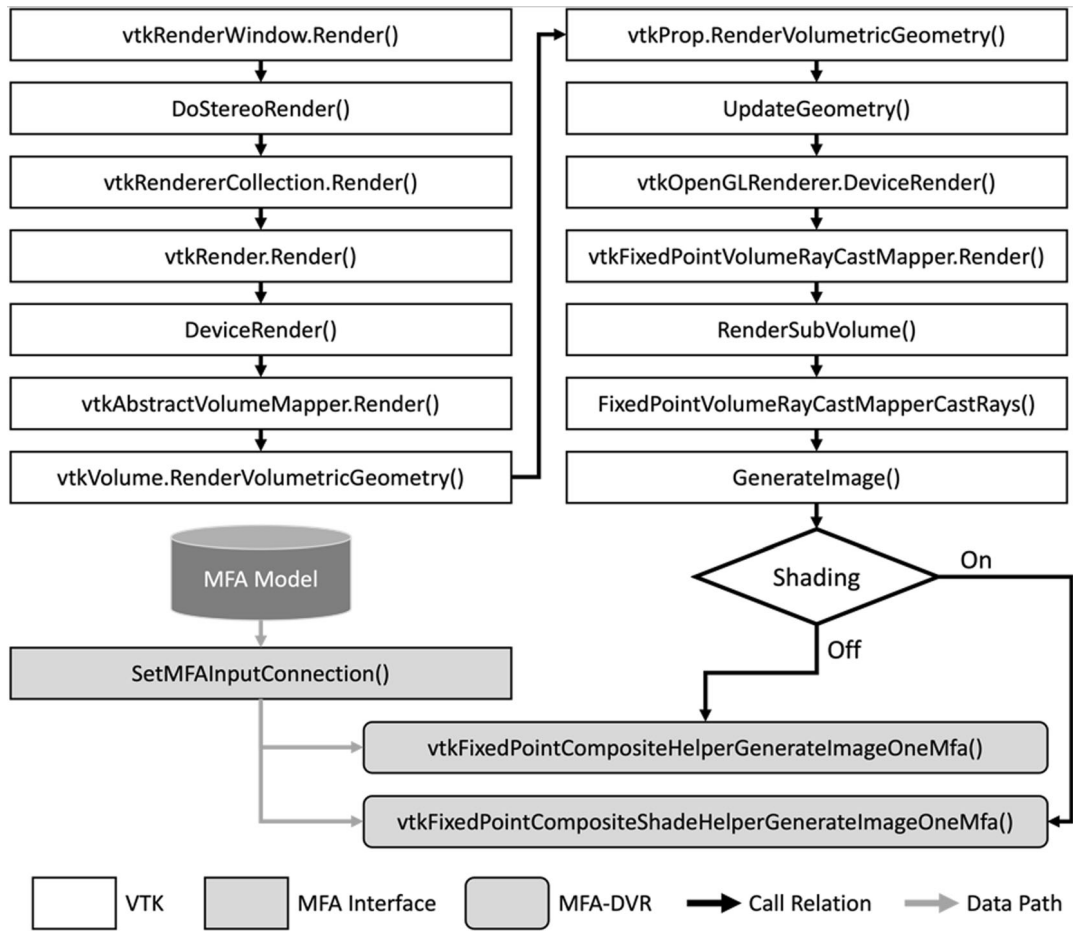

**Fig. 3** MFA-DVR call relation and data path through VTK

### 4.1 Dataset and experiment setup

#### 4.1.1 Synthetic datasets

In order to generate ground truth for quantitative evaluation, we use two synthetic functions, Gaussian Beam and Marschner–Lobb, which are multivariate analytical functions in 3D providing ground truth value and gradient anywhere in the domain. We also generate the corresponding discrete datasets from those functions as input to MFA-DVR and other discrete volume rendering algorithms.

*Gaussian Beam*, in its most basic form, is a multivariate radial basis function (RBF) in 3D, which characterizes a beam of electromagnetic radiation whose amplitude envelope in the transverse plane is given by a Gaussian function. The Gaussian Beam volume dataset is constructed by calculating the intensity of location at $x, y, z$ from:

$$F_{\text{GaussianBeam}}(x, y, z) = V_{\min} + (V_{\max} - V_{\min}) \cdot G\left(\frac{\sqrt{x^2 + y^2 + z^2}}{R}\right) \tag{1}$$

where

$$G(l) = e^{-\frac{(l-\mu)^2}{2\sigma^2}}$$

$V_{\min}$ and $V_{\max}$ define the lower and upper bounds of volume intensity; $G(l)$ is a Gaussian function, and $R$ is the maximum radius distance ($\sqrt{x^2 + y^2 + z^2} \leq R$). In our experiments, we use $V_{\min} = 0$, $V_{\max} = 255$, $\mu = 0$, $\sigma = 1/3$, and $R = \sqrt{3}$. The volume range of $x$, $y$, and $x$ is $[-1, 1]$.

*Marschner–Lobb* is a function originally proposed for the comparison of 3D resampling filters applied on the traditional Cartesian cubic lattice (Marschner and Lobb 1994). Due to the dynamics of its amplitude distribution across frequencies, it is challenging to reconstruct an accurate Marschner–Lobb signal from discrete samples, making it an effective benchmark for reconstruction quality. The Marschner–Lobb function used in this paper is defined as follows:

$$F_{\text{MarschnerLobb}}(x, y, z) = \frac{1 - \sin\left(\frac{\pi z}{2}\right) + \alpha \cdot \left(1 + \rho_r\left(\sqrt{x^2 + y^2}\right)\right)}{2(1 + \alpha)} \tag{2}$$

where

$$\rho_r(r) = \cos\left(2\pi \cdot f_M \cdot \cos\left(\frac{\pi r}{2}\right)\right)$$

We use $f_M = 6$ and $\alpha = 0.25$ to generate the discrete datasets.

#### 4.1.2 Real datasets

We also use real discrete datasets to demonstrate the rendering quality of MFA-DVR. Fuel and Nucleon are structured datasets of size $41^3$ and $64^3$, respectively, and represent examples of sparse datasets or situations of zooming in view on a small number of data samples from a large dataset. Interpolation artifacts are more obvious to observe on such datasets when off-grid interpolation is not accurate. Fuel is the simulation of fuel injection into a combustion chamber, while Nucleon is the simulation of the two-body distribution probability of a nucleon in an atomic nucleus. Aneurysm and Bonsai are structured datasets with the size of $512^3$ and $256^3$, respectively, and represent examples of larger-scale datasets. Aneurysm is the rotational angiography scan of a head with an aneurysm, while Bonsai is the CT scan of a bonsai tree. The NASA dataset is an unstructured dataset from NASA's Fun3D Retropropulsion simulation (FUN3D 2020), where the distribution and size of tetrahedras vary widely. The NASA dataset used in our experiment contains 27,736 samples.

*4.1.3 Experiment setup*

The hardware platform running the following experiments is a laptop with Intel Core i5 2.3 GHz quad-core CPU, Intel Iris Plus Graphics 655 and 16 GB system memory. The operating system is macOS 12.1. The VTK version used to implement MFA-DVR is 9.0. During ray traversal, for a fair comparison, we set the sample distance on each ray the same value for both MFA-DVR and traditional discrete volume rendering algorithms. This ensures that, for the same dataset, the total number of samples used for different volume rendering algorithms is the same. For a given dataset, we also use the same opacity and color transfer functions across all the volume rendering algorithms evaluated.

4.2 Rendering quality evaluation

*4.2.1 Quantitative evaluation via metrics*

To evaluate the rendering quality of MFA-DVR, we generate the ground truth volume rendering image by computing the value and gradient analytically from the synthetic functions for the samples on rays. Metrics of image quality include mean squared error (MSE), peak signal-to-noise ratio (PSNR), and structural similarity index (SSIM). Such metrics are computed over the resulting volume-rendered image of various algorithms compared with the ground truth volume-rendered image at the pixel level.

*Value accuracy evaluation* For various volume rendering algorithms, the accuracy of the values interpolated at ray sample points can be reflected through the quality of volume rendering image. We evaluate the accuracy of the value calculated from various 3D interpolation algorithms and our proposed MFA-DVR using a dataset containing multiple Gaussian beams with various scales. For only evaluating value accuracy, shading and gradient are disabled for this evaluation. The left figure of Fig. 4 shows the volume rendering of the multiple Gaussian beam datasets, where the center coordinates of all Gaussian beams share the same z value, and their x and y coordinates are aligned diagonally. The ramp function is used as the opacity transfer function to capture values across the entire range of the dataset. The color transfer function is constant red for its range. Four zoom levels are selected to show how those algorithms perform on value accuracy with various 3D resolutions. From zoom levels 1 to 4, as zooming in, each covers a Gaussian Beam with 3D resolutions of $64^3$, $32^3$, $16^3$ and $8^3$. The right figure array ($4 \times 8$) of Fig. 4 shows the volume rendering of each zoom level using different algorithms and the ground truth. The comparisons of rendering accuracy metrics are shown in Fig. 5. We observe that higher 3D resolution increases fidelity across all algorithms. However, as the zoom level increases, 3D samples covered by the view become sparser, and all algorithms suffer from accuracy degradation. High-order local filters like Tricubic and Catmull-Rom perform better than linear algorithms including Bunyk, Zsweep and PT. Across all zoom levels, MFA-DVR gives the best score on value accuracy compared with other algorithms. Figure 6 shows the visualization of pixel level rendering error distribution using different algorithms compared with the ground truth.

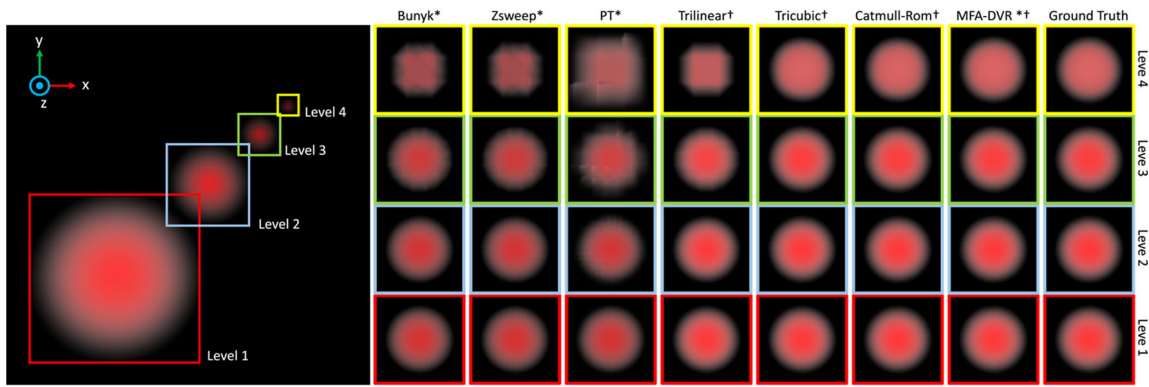

**Fig. 4** Volume rendering quality comparison under various zoom levels. Algorithms denoted with a star are for rendering unstructured data while algorithms with a dagger are for rendering structured data. MFA-DVR is capable of rendering both structured and unstructured data

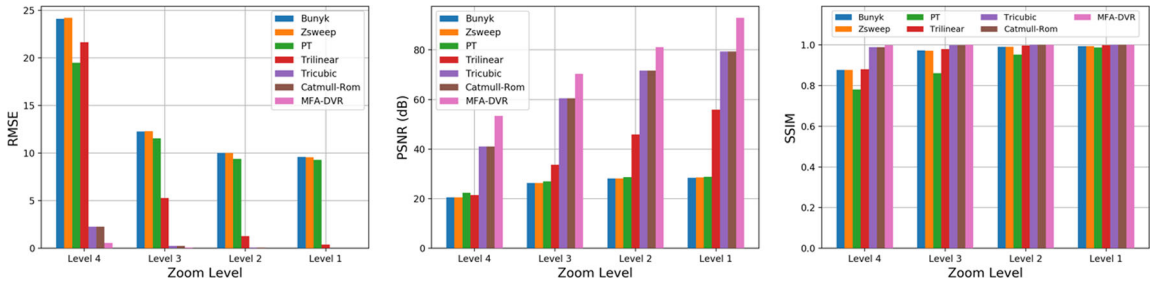

**Fig. 5** Quantitative evaluation of value accuracy using different algorithm under various zoom levels

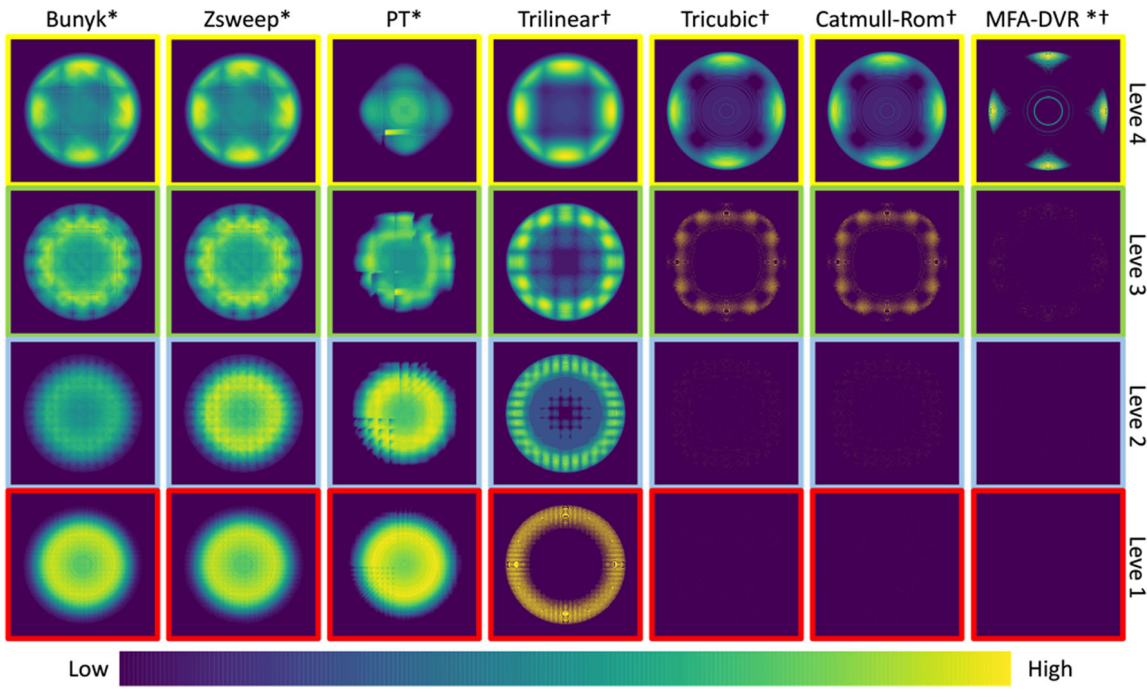

**Fig. 6** Visualization of pixel level rendering error distribution using different algorithms compared with the ground

*Gradient accuracy evaluation* Given the variations in gradient estimation methods (Correa et al. 2009) for unstructured datasets, here, we only evaluate the gradient accuracy of structured rendering algorithms for a fair comparison. Volume rendering with shading requires two steps, value query for mapping opacity and color through transfer functions, followed by gradient query for adding shading effects. In order to only evaluate gradient accuracy, we use the same ground truth analytical value across all the algorithms and then, calculated gradient values using respective algorithms. In this way, all the rendering will reflect only the gradient accuracy through shading. The Marschner–Lobb dataset is used to test the gradient accuracy due to its dynamic structure. A step function is used as the opacity transfer function in order to construct a surface for lighting interaction. The color transfer function is constant white for its range. Figure 7a shows the volume rendering results using the same ground truth value but different gradients calculated from various algorithms. Since the contour of the isosurface is defined only by the value, the contours of rendering images using all the algorithms are the same for they share the same ground truth value, which can be seen from the second row of Fig. 7a. However, on top of the same isosurface, different algorithms create different shading due to difference in the gradient calculation. This can be observed as ripples of specular lighting shown in the second and third rows of Fig. 7a when using trilinear, tricubic and Catmull-Rom local filters. MFA-DVR gives a much cleaner distribution of lighting which is close to the ground truth. The comparisons of error

**Table 1** Quantitative evaluation of gradient accuracy using different algorithms

| Metrics | Trilinear | Tricubic | Catmull-Rom | MFA-DVR |
|---|---|---|---|---|
| MSE | 601.01 | 1097.50 | 1099.48 | **196.30** |
| PSNR(dB) | 20.34 | 17.73 | 17.72 | **25.20** |
| SSIM | 0.76 | 0.79 | 0.79 | **0.93** |

Our method (MFA-DVR) gives the lowest MSE (Mean squared error, lower the better), highest PSNR (Peak signal-to-noise ratio, higher the better), and the highest SSIM (structural similarity index, higher the better) for the two tests (in bold) metrics on gradients are shown in Table 1. MFA-DVR also gives the best score on gradient accuracy compared with other algorithms.

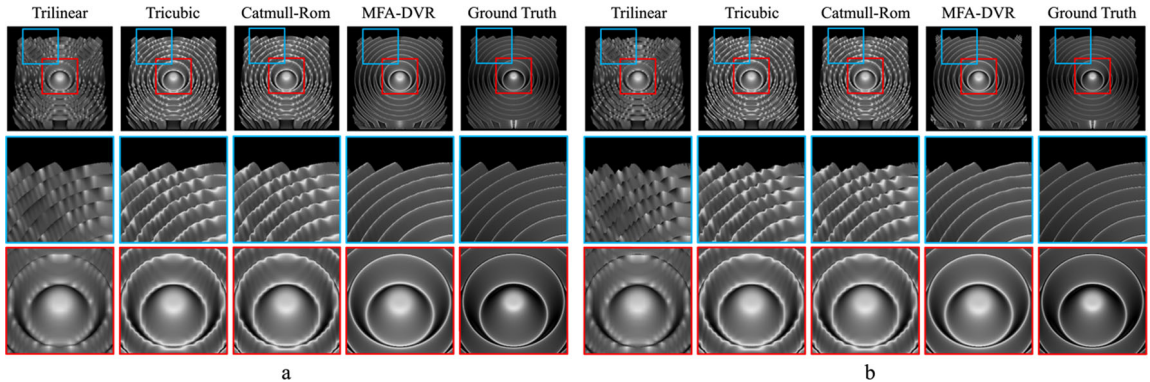

**Fig. 7** Volume rendering with shading for evaluating the value and gradient query accuracy using various algorithms

Next, we consider the overall accuracy using both value and gradient calculated from different algorithms. The rendering results and scores are shown in Fig. 7b and Table 2. In this case, the overall quality becomes worse with aggregated error from both value and gradient calculation compared to Table 1, where only gradient error is reflected. We can clearly see that imperfect value calculation introduces error on the surface reconstruction as shown in the second row of Fig. 7b. MFA-DVR still gives the best overall results for its higher accuracy on both value and gradient.

### 4.2.2 Qualitative evaluation via visualization

We also evaluate the rendering quality of MFA-DVR using real discrete datasets. Since the ground truth value and gradient are missing at the off-grid locations for discrete datasets, we evaluate the rendering quality via visualization.

*Structured dataset* Figure 8 shows the volume rendering results with shading on structured datasets using different algorithms. Trilinear produces aliasing artifacts on the lower edge of the ring in the Nucleon dataset and thin branches in the Bonsai dataset. It also tends to render small objects into diamond shapes as seen in the Fuel and Aneurysm datasets. The linear filter also generates star-shaped specular lighting artifacts on the surface of objects with sharp curvature. Although high-order filters like Tricubic and Catmull-Rom can mitigate such high-frequency artifacts on simple datasets like Nucleon and Fuel, they occasionally fail to reconstruct correct the structure or lighting of a complex dataset, like Aneurysm and Bonsai, due to overfitting inside the range of the local filter, especially near the boundaries of the dataset.

**Table 2** Quantitative evaluation of overall accuracy of both value and gradient using different algorithms

| Metrics | Trilinear | Tricubic | Catmull-Rom | MFA-DVR |
|---|---|---|---|---|
| MSE | 660.72 | 1174.36 | 1176.47 | **254.1** |
| PSNR(dB) | 19.93 | 17.43 | 17.43 | **24.08** |
| SSIM | 0.74 | 0.77 | 0.77 | **0.92** |

Our method (MFA-DVR) gives the lowest MSE (Mean squared error, lower the better), highest PSNR (Peak signal-to-noise ratio, higher the better), and the highest SSIM (structural similarity index, higher the better) for the two tests (in bold)

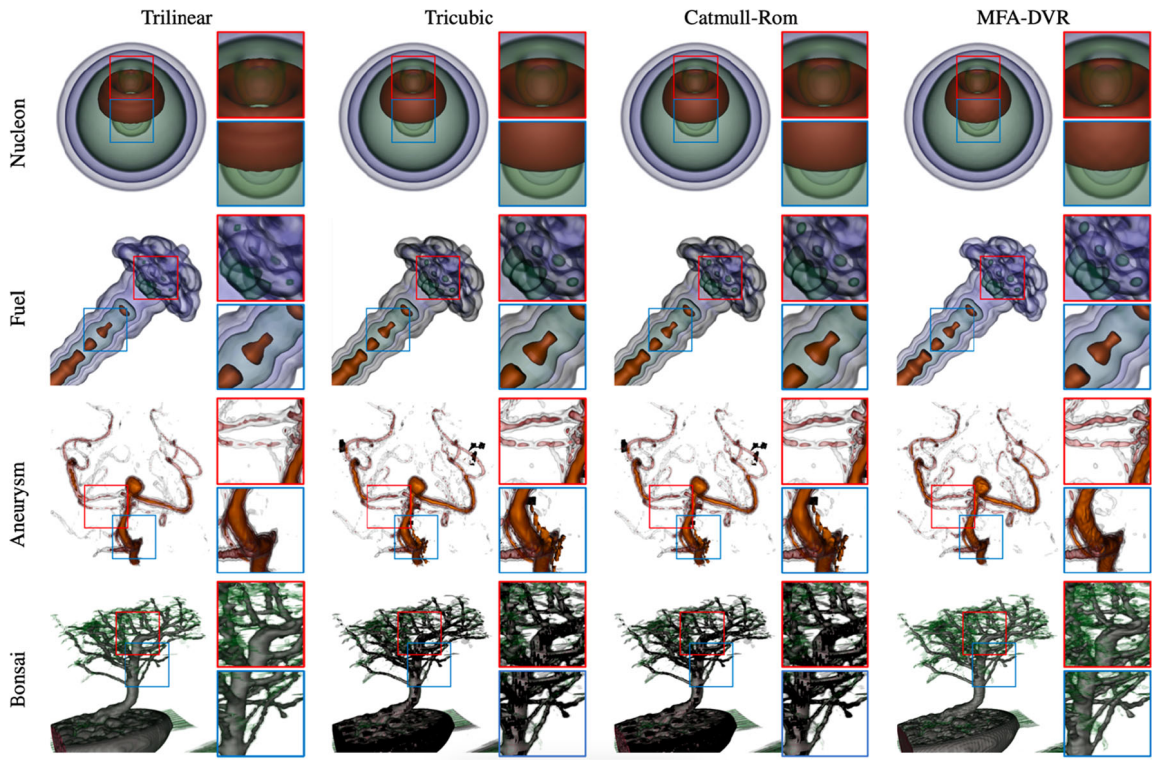

**Fig. 8** Volume rendering results on structured datasets using trilinear, tricubic, Catmull-Rom, and MFA-DVR

MFA-DVR, as a high-order approximation, is not only smoother than linear filters but also capable of reconstructing more accurate structure and lighting for both sparse and dense datasets.

*Unstructured dataset* Figure 9 shows the volume rendering results on the unstructured NASA dataset. Shading is turned off to clearly visualize the rendering quality. Similar advantages of using MFA-DVR are observed, as MFA-DVR eliminates high-frequency artifacts with smooth structure compared with Bunyk and Zsweep. Results of Projected Tetrahedra produce blurry results that miss detailed structures.

### 4.2.3 MFA Parameter study on rendering quality

In this section, we investigate how the key MFA parameters, number of control points and polynomial degree, affect the final volume rendering quality. We use the Marschner–Lobb dataset of size $256^3$ for the evaluation and compute the metric scores of MSE, PSNR, and SSIM.

The top row of Fig. 10 shows how the number of control points affects each rendering quality metric across different polynomial degrees. We observe the trend of quality degradation while using fewer control points. However, when using a degree higher than 1, this trend does not start to dominate until the number of control points is set too low. In other words, with a higher degree, fewer number of control points are needed to achieve a similar level of rendering quality. Using a degree of one generally gives suboptimal quality because the MFA becomes a linear model, making MFA-DVR generates similar artifacts to trilinear interpolation. Using a very small or very large number of control points might be detrimental to the final rendering quality as shown in the second row of Fig. 10. Using a very small number of control points will not be enough to capture the dynamics of the input dataset, as a result, no matter what the polynomial degree MFA encoder uses, the final rendering will not be satisfactory. On the other hand, if too many control points are used, more than enough control points are used to capture the dynamics for the input dataset, giving MFA a hard time searching for the optimal placement of that many control points for a given constraint on encoding time complexity. The bottom row of Fig. 10 also shows how the polynomial degree affects those metrics using a different number of control points. Using a higher degree normally results in a better render quality, except for extreme cases where the number of control points is too large or too small. When the number of control points is close to the size of the input discrete dataset, using a high polynomial degree generates Runge's phenomenon, which is detrimental to rendering quality because of the oscillations due to overfitting at data boundaries. Otherwise, using a higher degree normally gives a better SSIM score meaning a more faithful structural reconstruction. This is because high-frequency noise is smoothed. In general, using a polynomial degree greater than one will help to generate higher-quality rendering results since most scientific data are nonlinear. However, the optimal order depends on the dynamics of the specific data. MFA-DVR is elastic on the number of control points which enables MFA-DVR to give relatively high rendering quality without using many control points. For example, using one-half the number of control points as the original data points in each dimension, and a degree of 2 or 3 works well for the Marschner–Lobb dataset.

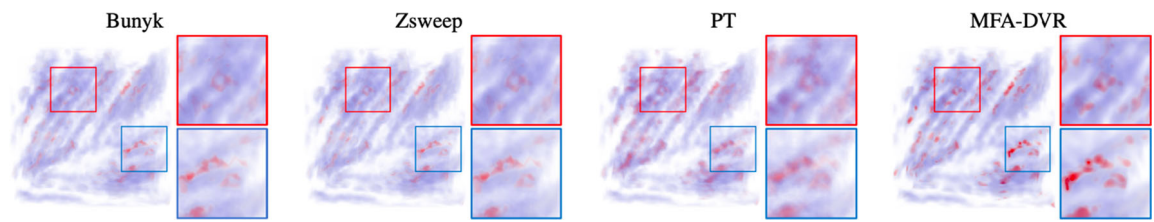

**Fig. 9** Volume rendering results on unstructured NASA data using Bunyk, Zsweep, Projected Tetrahedra (PT), and MFA-DVR

### 4.3 Performance evaluation

Volume rendering performance on structured and unstructured datasets is evaluated using test datasets of various sizes. We construct structured datasets with various sizes to discover how those volume rendering algorithms performance and scale. The number of control points is the main MFA parameter that determines the compression ratio of the MFA model. Specifically, the size of the MFA model is proportional to the number of control points used for encoding, while the polynomial degree has no influence on the size of the MFA model. If the number of the control points used for encoding the MFA model is the same as the input sample size of the input volumetric data on each of the x, y, and z directions, the encoded MFA model size will also be the same size of the input volumetric data. This is how we generate the MFA model with the sizes matching the input volumetric data. The MFA model has infinite resolution regardless of the element size and shape of the unstructured dataset. Since only the rendering speed is investigated, we generate unstructured test datasets from the same structured test datasets by converting regular grid into unstructured cells so that all algorithms use the same input data. In this work, we run all algorithms, except for PT, on the same CPU platform for a fair comparison. The performance measured will reflect the real computational complexity of each algorithm although they all can be accelerated by multi-core architectures. We also investigate a cross-platform comparison of performance between MFA-DVR on CPU and PT on GPU for tetrahedral cells that are widely adopted in GPU acceleration techniques. The test dataset is the single Gaussian Beam synthetic dataset with various sizes for rendering an image of 1000 × 1000 resolution. For MFA-DVR, the number of control points is set to the data size for each spatial dimension of the input volume, and the polynomial degree is set to 2 for higher-order interpolation.

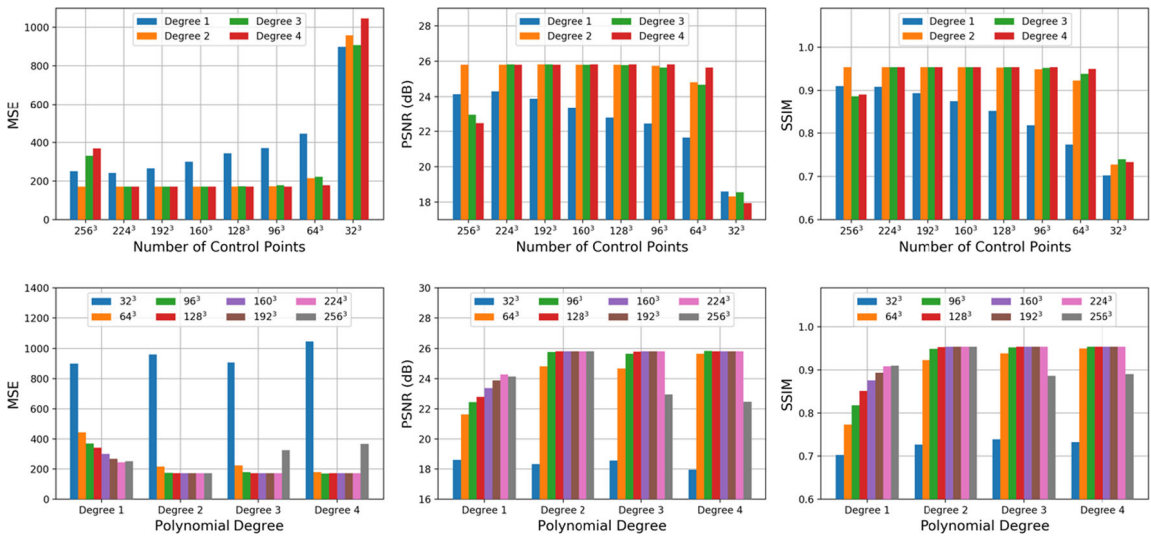

**Fig. 10** MFA-DVR rendering quality using different number of control points and polynomial degree

### 4.3.1 Structured volume rendering

The left figure of Fig. 11 shows the rendering time of different structured volume rendering algorithms and MFA-DVR on the testing dataset with various sizes. Trilinear is the fastest for rendering structured data. The experiment shows the MFA-DVR is about eight times slower than trilinear due to the overhead of querying the MFA model. However, compared with other higher-order interpolation algorithms, tricubic and Catmull-Rom, MFA-DVR performs much faster as a high-order approximation. This is because MFA-DVR only executes querying through analytical computation with fitting parameters already calculated at priori rather than doing expensive fitting on the fly. This approves that MFA-DVR performs well as a post hoc visualization methods. On the other hand, the time complexity of MFA-DVR is almost independent of the growth of the dataset, making MFA-DVR scales well when rendering large-scale dataset.

### 4.3.2 Unstructured volume rendering

The middle figure of Fig. 11 shows the comparison of rendering time between MFA-DVR and other unstructured volume rendering algorithms running on a CPU. Since the MFA model has a uniform representation for both structured and unstructured data, MFA-DVR avoids complicated searching and projecting operations on large unstructured meshes. As a result, under the same platform, MFA-DVR not only renders faster but also scales better on unstructured datasets than Bunyk and Zsweep. The right figure of Fig. 11 shows the rendering time comparison between MFA-DVR on CPU and Projected Tetrahedra on GPU. Although projection-based unstructured volume rendering algorithms, like Projected Tetrahedra, utilize the GPU to shorten the rendering time, their complexity with respect to data size has superlinear growth, while MFA-DVR scales linearly with data size. This is because the MFA decoding is only doing sample size number of sum operations on each of the three dimensions. Although computational benefits brought by a newer generation of GPU can speed up projection-based unstructured volume rendering algorithms, MFA-DVR shows potential for efficiently rendering large-scale datasets.

### 4.3.3 MFA Parameter study on performance

We also investigate how the key MFA parameters affect the querying time, which determines the overall rendering time of MFA-DVR. The input dataset used is Gaussian Beam of size $256^3$. We measure how long it takes for the querying function to return the value or gradient from the MFA model for one sample location on ray. The left column of Fig. 12 shows how the number of control points and polynomial degree influence the value query time. We see that the growth of the number of control points results in longer value querying time, but such an effect is linear and not as significant as the growth due to polynomial degree. A similar pattern can be seen in gradient querying time, as the middle column of Fig. 12 shows. The overall query time for value and gradient together using different values of MFA parameters is shown in the right column of Fig. 12.

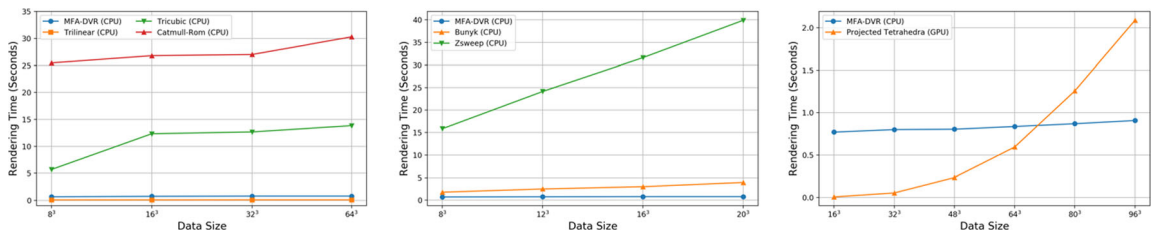

**Fig. 11** Quantitative evaluation of rendering time using different algorithms

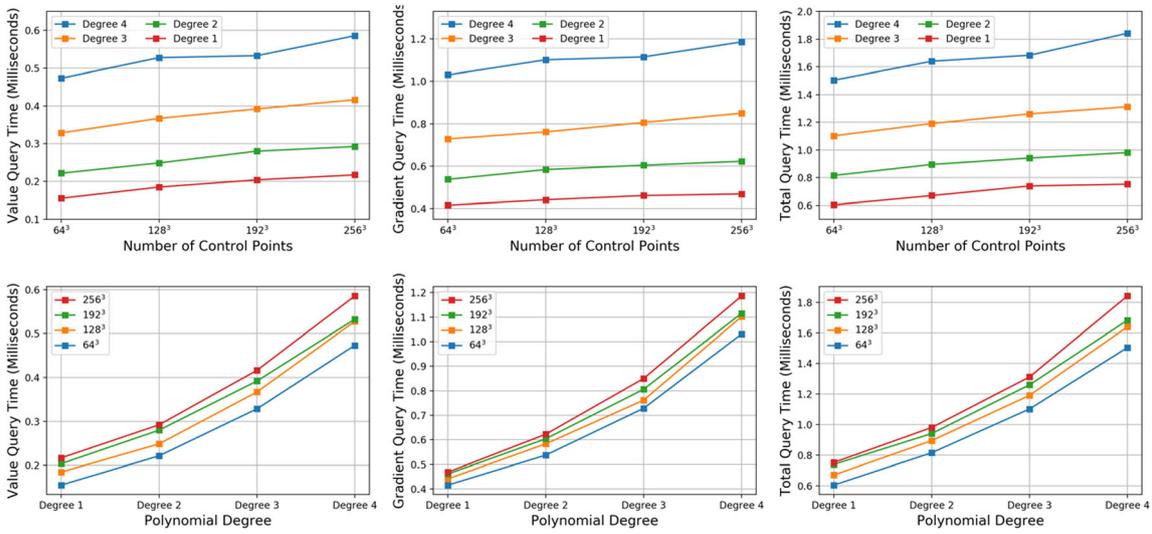

**Fig. 12** MFA-DVR value and gradient query performance using different number of control points and polynomial degree

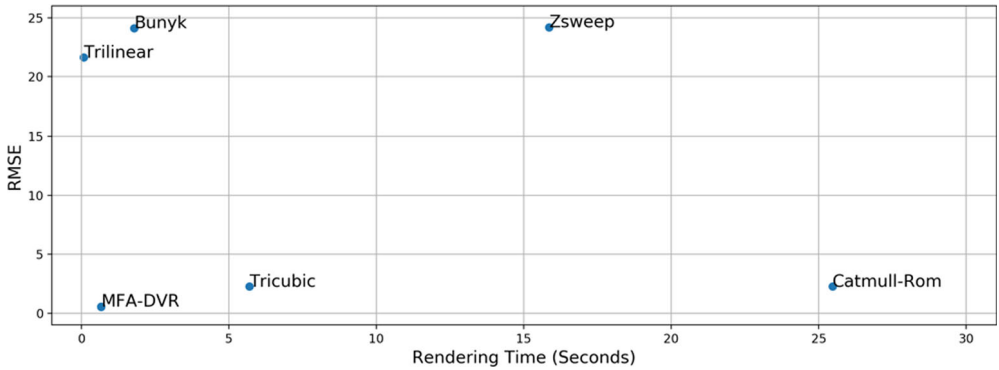

**Fig. 13** Overall quality and performance evaluation of MFA-DVR and other volume rendering algorithms

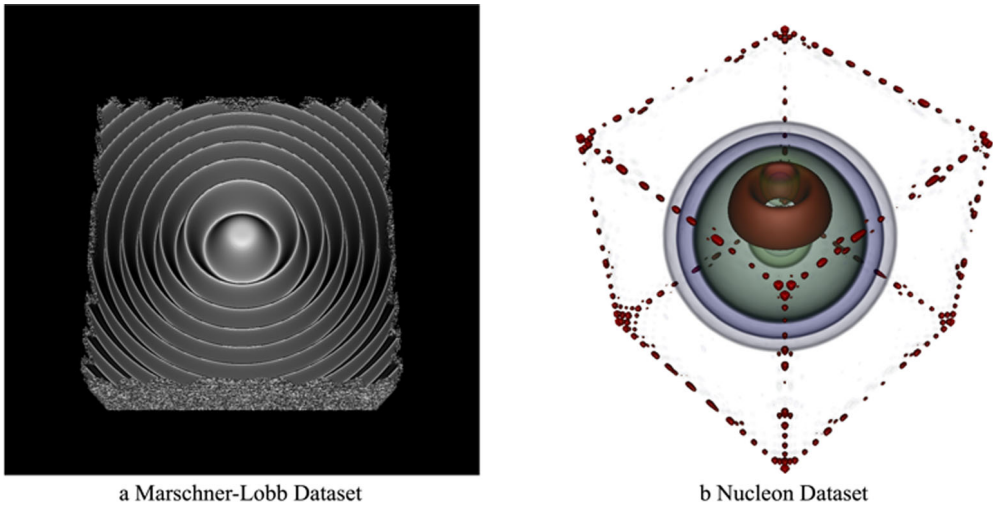

**Fig. 14** Rendering artifacts due to Runge's phenomenon when using suboptimal high polynomial degree order for specific dataset

### 4.4 Quality and performance

Figure 13 shows the relative positions of all algorithms running on CPU considered in this study on accuracy (RMSE) and performance (rendering time) dimensions. Trilinear is the fastest with bad quality while high-order local filters give better quality with slow performance. Unstructured algorithms give worse quality with relatively slow performance. MFA-DVR gives the best quality with the fastest rendering time except for only the trilinear method.

## 5 Discussion

As discussed in Sect. 4.2.3, for complex datasets, using a large number of control points with a high polynomial degree will give inaccurate fitted values at data boundaries due to Runge's phenomenon. Figure 14a shows the volume rendering result of the MFA model encoded from the Marschner–Lobb dataset using the same number of control points as data size with a polynomial degree of 4. Figure 14b shows a similar result for Nucleon dataset using the MFA model encoded with a large number of control points and a high polynomial degree of 3. Both results have noticeable errors around the boundary of the volume cube. For complicated data, encoding an MFA model with a large number of control points with a high encoding degree should be avoided.

This work is an initial exploration of utilizing the MFA model in data visualization applications. Although we use volume rendering as an example to show the potential of the MFA model, we believe other data visualization algorithms can also leverage the MFA to query information from the model on demand to improve the visualization quality and performance. Our current VTK implementation of MFA-DVR only supports a single light source for volume rendering with shading; multiple light sources can be added in the future. We are also interested in implementing the MFA-DVR algorithm on the GPU to further speed up its performance. Other MFA features like data compression and noise reduction can also be explored in future work.

## 6 Conclusion

In this work, we propose MFA-DVR, the first volume rendering pipeline for directly rendering MFA model. MFA-DVR leverages MFA features to improve multiple aspects of volume rendering. We design the pipeline of directly rendering an encoded MFA model with optimized interfaces querying value and gradient from MFA model on arbitrary locations with high accuracy and efficiency. MFA-DVR mitigates linear interpolation artifacts, performs better on accuracy and performance than classic high-order filters, and speeds up volume rendering on unstructured datasets. We validate the efficacy of MFA-DVR quantitatively and qualitatively through experiments and evaluations. Key MFA parameters used for encoding the MFA model are also investigated to see how they affect the rendering quality and performance. Our implementation makes it easy to try MFA-DVR on existing visualization codebase utilizing VTK rendering pipeline.

The original MFA is designed as a general model for high-dimensional scientific data. This work is an initial exploration of utilizing the MFA model in volume visualization applications. Although we use volume rendering as an example to show the potential of the MFA model, we think other data visualization algorithms can also leverage our design of the DVR pipeline and optimization to improve their quality and performance. We are interested in implementing the MFA-DVR algorithm on the GPU to further speed up its performance.

**Acknowledgements** This work is supported by Advanced Scientific Computing Research, Office of Science, U.S. Department of Energy, under Contracts DE-AC02-06CH11357, program manager Margaret Lentz. The authors gratefully acknowledge the assistance and support of Berk Geveci and the entire Kitware team for their help in understanding and modifying the VTK rendering pipeline.